\documentclass{ws-ijmpa}

\newcommand {\beq}{\begin{equation}}
\newcommand {\eeq}{\end{equation}}
\newcommand {\bea}{\begin{eqnarray}}
\newcommand {\eea}{\end{eqnarray}}

\newcommand {\K}{{\bf K}}
\newcommand {\I}{{\bf I}}

\newcommand {\m}{\mu}

\newcommand {\pl}{\partial}

\newcommand {\al}{\alpha}
\newcommand {\be}{\beta}

\newcommand {\La}{\Lambda}

\newcommand {\om}{\omega}

\newcommand {\e} {\mbox{e}}

\newcommand {\del}  {\delta}

\newcommand {\half}{ {\frac{1}{2}} }

\newcommand {\Acal}{{\cal A}}

\newcommand {\Dcal}{{\cal D}}

\def\overleftarrow#1{\vbox{\ialign{##\crcr
 $\leftarrow$\crcr\noalign{\kern-1pt\nointerlineskip}
 $\hfil\displaystyle{#1}\hfil$\crcr}}}

\newcommand {\ptil}{{\tilde p}}

\newcommand {\intp} {{\int \frac{d^4p}{(2\pi)^4}}}

\newcommand {\change} {\leftrightarrow}

\newcommand {\com}  {{\quad ,}}
\newcommand {\q}    {\quad}

\newcommand {\Pla} {\frac{{\tilde p}}{\omega}}

\newcommand {\Tev} {\frac{{\tilde p}}{T}}

\begin{document}

\markboth{SHOICHI ICHINOSE}
{
Casimir Energy of 5D Electro-Magnetism and  
Sphere Lattice Regularization
}

%
\catchline{}{}{}{}{}
%

\title{
Casimir Energy of 5D Electro-Magnetism and  
Sphere Lattice Regularization
}

\author{SHOICHI ICHINOSE}

\address{
Laboratory of Physics, School of Food and Nutritional Sciences, 
Univ. of Shizuoka\\
Yada 52-1, Shizuoka 422-8526, Japan
\\
ichinose@u-shizuoka-ken.ac.jp}

\maketitle

\begin{history}
\received{Day Month Year}
\revised{Day Month Year}
\end{history}

\begin{abstract}
Casimir energy is calculated in the 5D warped system.
It is compared with the flat one. The position/
momentum propagator is exploited. A new regularization, 
called {\it sphere lattice regularization}, is introduced. 
It is a direct realization of the geometrical interpretation 
of the renormalization group.  
The regularized configuration is closed-string like. 
We do {\it not} take the KK-expansion approach. Instead 
the P/M propagator is exploited, 
combined with the heat-kernel method. All expressions
are closed-form (not KK-expanded form). Rigorous quantities 
are only treated (non-perturbative treatment). 
The properly regularized form of Casimir energy, is expressed in the closed form. We
numerically evaluate its $\La$(4D UV-cutoff), $\om$(5D bulk curvature, 
warpedness parameter)
and $T$(extra space IR parameter) dependence.
\end{abstract}


Casimir energy is the free part of the vacuum energy. 
It depends only on the macro (boundary) parameters. 
It is the macroscopic quantum effect. 
In the recent strong interest in the brane models or
the bulk-boundary theories, 
the subject is quite important. 
The higher dimensional Casimir energy was examined by 
Appelquist and Chodos\cite{AC83}. 
They considered the flat geometry of $S^1\times{\cal M}_4$. 
It has been, in the recent standpoint, re-examined
\cite{SI0801}. Here we examine the warped case. 

In the closed form, $E_{Cas}$ of 5D electro-magnetism is expressed as 
$E_{Cas}(\om,T)=\half E^-_{Cas} +2E^+_{Cas},\mbox{(1)}$\ 
$E^\mp_{Cas}(\om,T)
=\intp\int_{1/\om}^{1/T}dz~s(z)\int_{p^2}^\infty\{G_k^\mp (z,z)\}dk^2$ 
$\equiv \intp\int_{1/\om}^{1/T}dzF^\mp(\ptil,z), s(z)=1/(\om z)^3
,$(2)\ 
%
where the P/M propagators are 
$
G_p^\mp(z,z')=\mp\frac{\om^3}{2}z^2{z'}^2
\frac{\{\I_0(\Pla)\K_0(\ptil z)\mp\K_0(\Pla)\I_0(\ptil z)\}  
      \{\I_0(\Tev)\K_0(\ptil z')\mp\K_0(\Tev)\I_0(\ptil z')\}
     }{\I_0(\Tev)\K_0(\Pla)-\K_0(\Tev)\I_0(\Pla)}\ ,
\ptil\equiv\sqrt{p^2}\com\q p^2\geq 0\ (\mbox{space-like})
$. 
The integral region of the equation (2) 
is displayed in Fig.1. 
In the figure, we introduce the UV and IR regularization cut-offs, 
$\m=\La T/\om\leq\ptil\leq\La$, $1/\om\leq z\leq 1/T$.
From a close numerical analysis of ($\ptil,z$)-integral (2), 
we have confirmed\q
$E^{\La,-}_{Cas}(\om,T)=(1/8\pi^2)\left\{ -0.0250 \La^5/T\right\}
$. \q
This is the same result as the flat case\cite{SI0801} (periodicity $2l$)
by the replacement : $1/5T \change l$.   
The $\La^5$-divergence shows the notorious problem
of the higher dimensional theories. 
In spite of all efforts of the past literature, 
we have not succeeded 
in defining the higher-dimensional theories.

\begin{eqnarray}
\begin{array}{cc}
\begin{array}{c}
\mbox{\psfig{file=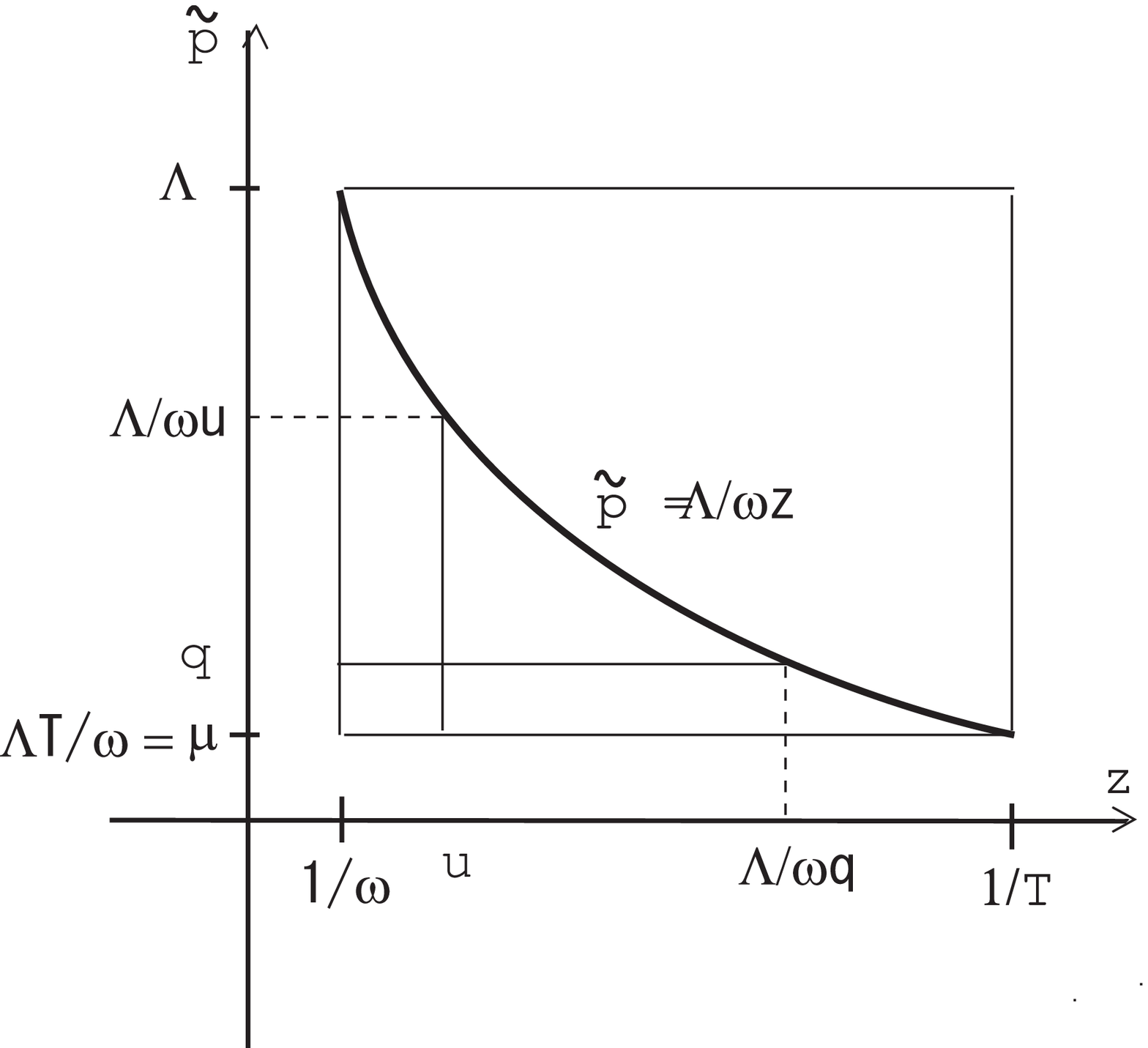,height=35mm}}\\
\parbox{50mm}{Fig.1\ Space (z,${\tilde p}$) for integral}\\
\mbox{\psfig{file=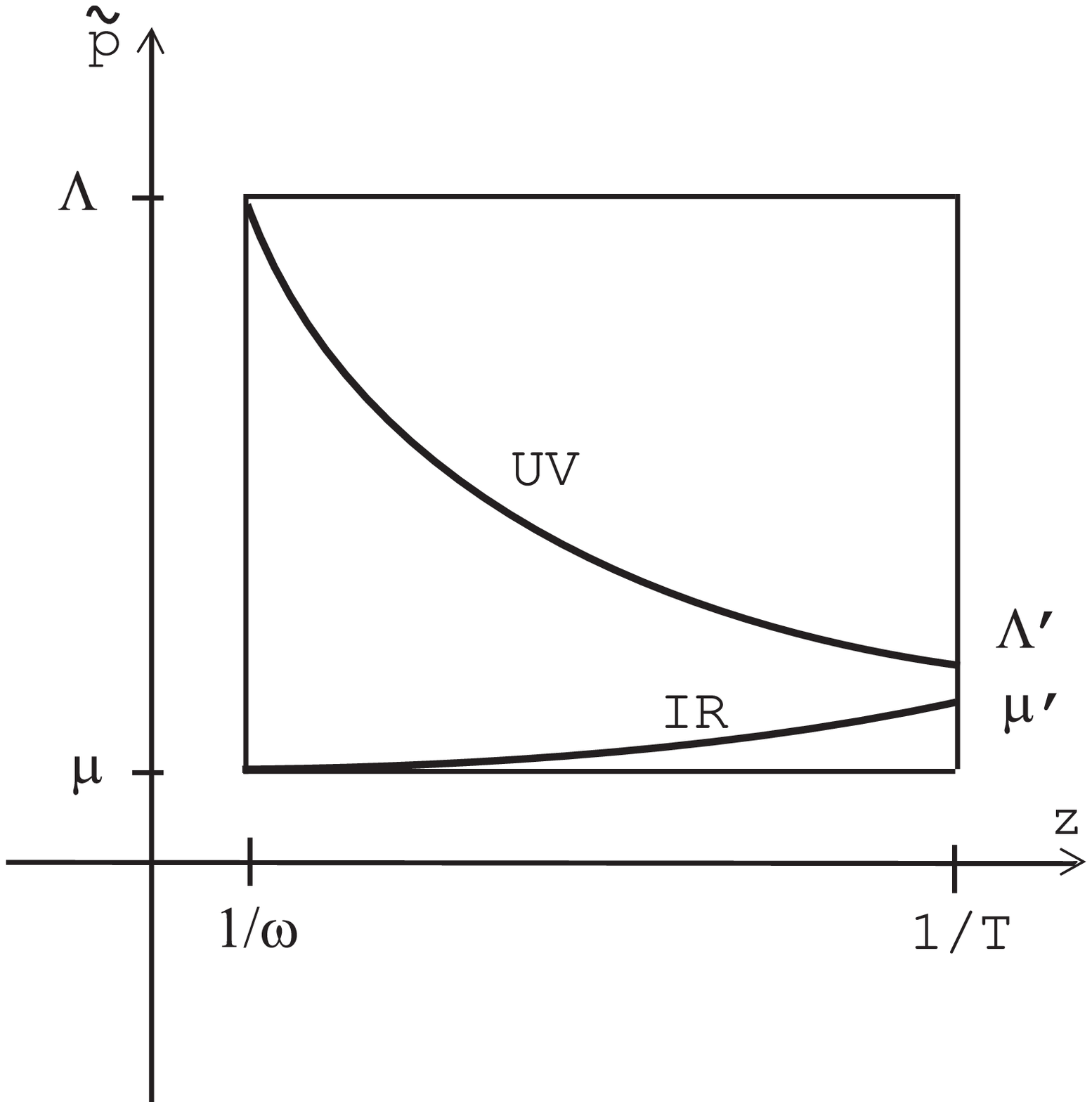,height=35mm}}\\
\parbox{50mm}{Fig.2\ Proposed int. space (z,$\ptil$)}
\end{array} &
\begin{array}{c}
\mbox{\psfig{file=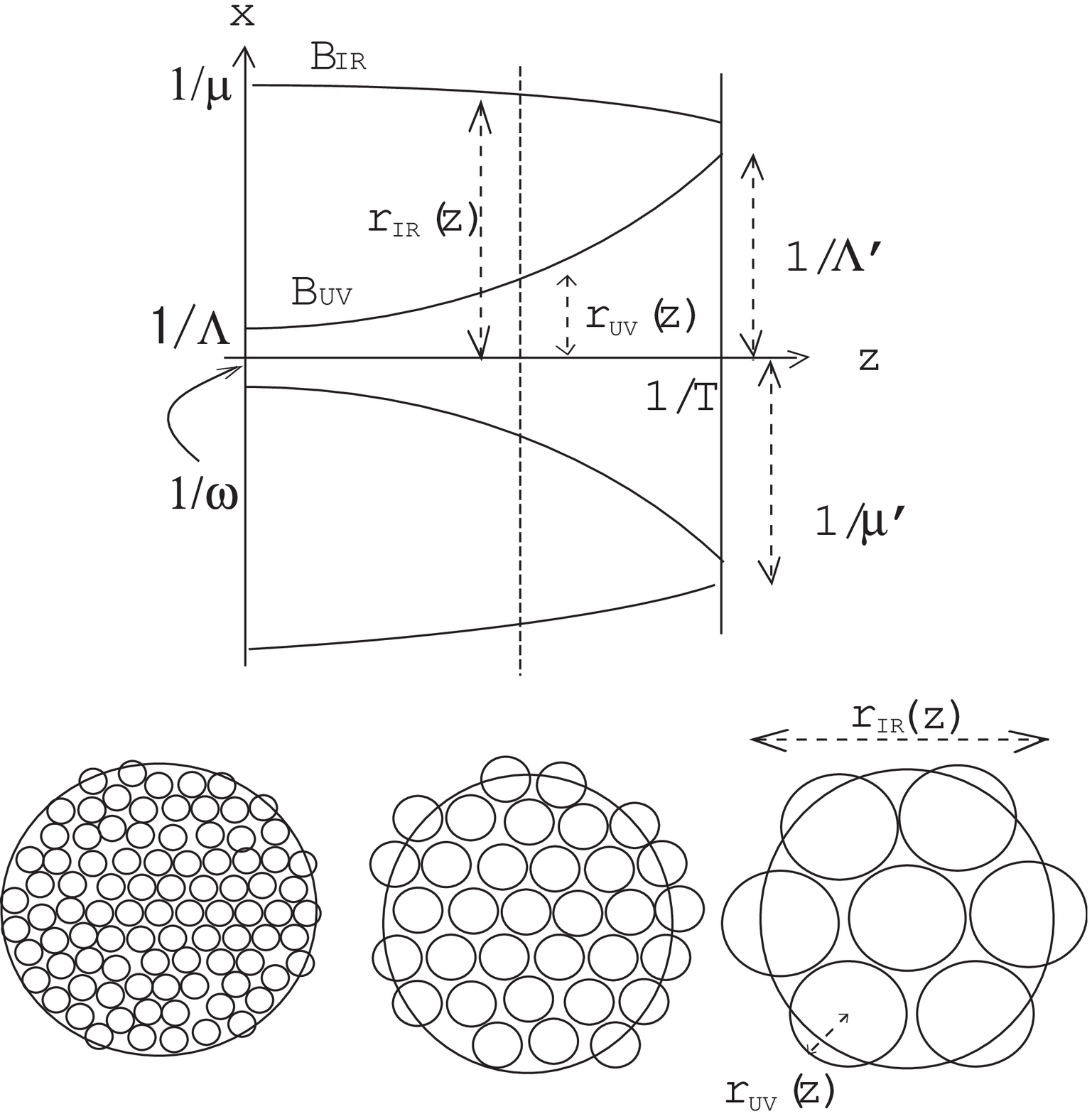,height=7cm}}\\
\parbox{90mm}{Fig.3\ Regularization Surface $B_{IR}$ and $B_{UV}$ 
in the 5D coordinate space $(x^\m,z)$, Flow of Coarse Graining 
(Renormalization) and Sphere Lattice Regularization.}
\end{array}
\end{array}
\nonumber
\end{eqnarray}

The divergence 
causes problems. 
The famous example is 
the divergent cosmological constant in the gravity-involving theories.
\cite{AC83} 
We notice here we can avoid the divergence problem if we find a way to
{\it legitimately restrict the integral region in ($\ptil,z$)-space}. One 
proposal of this was presented by Randall and Schwartz\cite{RS01}. They introduced
the position-dependent cut-off,\ $\mu=\La T/\om <\ptil <\La/\om u\ ,\ u\in [1/\om,1/T]$\ , 
for the 4D-momentum integral in the "brane" located at $z=u$. (See Fig.1)
The total integral region is the lower part of the hyperbolic curve $\ptil=\La/\om z$. 
(They succeeded in obtaining the {\it finite} $\be$-function in the 5D warped vector
model.) The numerical result says
$E^{-RS}_{Cas}(\om,T)=(1/8\pi^2)\left\{ -0.0167 \La^5/\om\right\}$. 
Compared with the flat case ($E^{RS}_{Cas}(l)=(5/8\pi^2)\left\{ -0.0178 \La^4\right\}$), 
the divergence situation does {\it not} improve unless we take $\La\propto \om$. 
Although they claim the holography is behind the procedure, 
the legitimateness of the restriction looks less obvious. We proposed
an alternate one \cite{IM0703}
and state here the legitimate explanation within the 5D QFT. 

On the "3-brane" at $z=1/\om$, we introduce the IR-cutoff $\mu$ and 
the UV-cutoff $\La$\ ($\mu\ll\La$). See Fig.2.  This is legitimate in the sense that we
usually do this procedure in the 4D {\it renormalizable} theories. On the
"3-brane" at $z=1/T$, we introduce another set of IR and UV-cutoffs, 
$\mu'$ and $\La'$. 
We consider 
the case: $\mu'\leq\La',\ \mu\sim\mu', \La'\ll\La$. This case leads us to
introduce the renormalization flow. (See the later explanation of Fig.3.)
We claim here,  
as for the "3-brane" located at each point $z$ ($1/\om<z<1/T$), the regularization 
parameters are determined by the {\it minimal area principle}. 
To explain it, we depict the regularization configuration of Fig.2 in the 5D coordinate space ($x^\m,z$) 
in Fig.3. 
The 5D volume region bounded by $B_{UV}$ and $B_{IR}$ is the integral region 
of the Casimir energy $E_{Cas}$. 
The forms of $r_{UV}(z)$ and $r_{IR}(z)$ can be
determined by the {\it minimal area principle}.\ 
$
\del (\mbox{Surface Area})=0,$
$3+\frac{4}{z}r'r-\frac{r''r}{{r'}^2+1}=0,$
$r'\equiv\frac{dr}{dz},$
$r''\equiv\frac{d^2r}{dz^2},$
$1/\om\leq z\leq 1/T.$\ (3)\ 
We have numerically confirmed the existence of appropriate geodesic 
curves for the 'inverse' renormalization flow. 

Instead of restricting the integral region, we have another approach to 
suppress UV and IR divergences. We introduce a {\it weight function} $W(\ptil,z)$.\q 
$
E^{\mp~W}_{Cas}(\om,T)\equiv\intp\int_{1/\om}^{1/T}dz~ W(\ptil,z)F^\mp (\ptil,z)
.\ $(4)\q
As the examples of $W$, we present $\e^{-\ptil^2/2\om^2-z^2T^2/2}\equiv W_1(\ptil,z)$({elliptic suppression})
and $\e^{-\ptil^2 z^2T^2/2}\equiv W_3(\ptil,z)$ ({hyperbolic suppression}). 
We have evaluated the divergence behavior of $E^W_{Cas}$ by 
numerically performing the $(\ptil,z)$-integral (4) 
for the rectangle region of Fig.1.   
$
E^{-W3}_{Cas}/\La T^{-1}=(1/8\pi^2)\{-2.61~10^{-2}\om\La^3-4.59~10^{-7}\om\La^3\ln\La\},\ 
E^{-W1}_{Cas}/\La T^{-1}=(1/8\pi^2)\{-0.312\om^4-1.06~10^{-2}\om^4\ln\La\} 
.\ $(5)\q 
$W_3$ corresponds to the restriction approach by Randall-Schwartz, but the above 
result does not coincide with theirs $E^{-RS}_{Cas}(\om,T)$. 
Its divergence-suppression is insufficient. 
$W_1$ gives, after normalization by the factor $\La/T$, the desired 
log-divergence only. In this case, the Casimir energy is {\it finitely} obtained by 
the {\it renormalization of the warp factor $\om$}.   
$
-\al \om^4(1-4c \ln(\La/\om))=-\al {\om'}^4,\ $
$
\frac{\pl}{\pl (\ln \La)}\ln\frac{\om'}{\om}=-c~(\mbox{anomalous dimension})
.\ $(6)\q 
(For $W_1$, $\al=0.312/8\pi^2, c=-0.849~10^{-2}$.)
Fig.3 shows the renormalization flow. For interacting theories, such as 
5D YM theories, 
the scaling of the renormalized coupling $g(z)$ is given by\ \ 
$
\be
=-\frac{1}{3}\frac{1}{g}\frac{\pl g}{\pl z}/{\frac{\pl}{\pl z}\ln r(z)},
\ $(7)\q 
where $g(z)$ is a renormalized coupling at $z$ and $r(z)$ is an appropriate geodesic.  

Finally we comment on the meaning of the weight function $W(\ptil,z)$.  
We can define it by {\it requiring} that the dominant contribution to $E^W_{Cas}$(4), 
which is obtained by the steepest-descend method to (4), 
{\it coincides} with the geodesic curve, which is obtained by the minimal area principle 
for the surface in the bulk(3).
For the purpose, we are naturally led to the following definition of the Casimir energy.
$E_{Cas}=\int d^4\al\int_{p^a(1/\om)=p^a(1/T)=\al^a}\Dcal^4p(z)F(\ptil,z)\exp\{\Acal\},$
where the '4D area action' $\Acal$, with Euclidean time $z$, is defined by 
the induced metric $K_{ab}(x)=(1/\om^2z^2)(\del_{ab}+x^ax^b/(rr')^2)$
on the surface $\sum_{a=1}^4(x^a)^2=r(z)^2=\ptil(z)^{-2}$ as 
$\Acal=\int\sqrt{K_{ab}}d^4x=\int_{1/\om}^{1/T}\sqrt{\ptil^{'2}/\ptil^4+1}/\om^2z^4\ptil^3~dz$. 
The energy distribution operator $F(\ptil,z)$ is defined by the ordinary field quantization (2). 
The above proposal means the 5D space $(p^a=(x^a)^{-1},z)$ is {\it quantized} with the action $\Acal[\ptil,\ptil',z]$ and the 
Euclidean time $z$. 

The absence of the proper definition of the higher dimensional QFT 
has been hindering the calculation of 5D Casimir energy so far. 
In this paper we have introduced a new definition and demonstrated a finite value.

\end{document}